\documentclass[superscriptaddress,twocolumn,showpacs,aps]{revtex4}

\usepackage{graphicx}% Include figure files
\usepackage{dcolumn}% Align table columns on decimal point
\usepackage{bm}% bold math

\begin{document}

\preprint{}

\title{
Quasi-elastic barrier distribution as a tool for 
investigating unstable nuclei}

\author{K. Hagino}
\affiliation{
Department of Physics, Tohoku University, Sendai 980-8578, Japan}

\author{N. Rowley}
\affiliation{
Institut de Recherches Subatomiques,
UMR7500, IN2P3-CNRS/Universtit\'e Louis Pasteur, BP28, F-67037
Strasbourg Cedex 2, France}

\date{\today}% It is always \today, today,

\begin{abstract}
The method of fusion barrier distribution has been 
widely used to interpret the effect of nuclear structure 
on heavy-ion fusion reactions around the Coulomb barrier. 
We discuss a similar, but less well known, barrier distribution 
extracted from large-angle quasi-elastic scattering. 
We argue that this method has several advantages over the fusion 
barrier distribution, and 
offers an interesting tool for investigating unstable nuclei.
\end{abstract}


\maketitle

\section{Introduction}

It has been well recognized that 
heavy-ion collisions at energies around the Coulomb barrier 
are strongly affected by the internal structure of 
colliding nuclei \cite{DHRS98,BT98}. 
The couplings of the 
relative motion to the 
intrinsic degrees of freedom (such as collective
inelastic excitations of the colliding nuclei and/or transfer
processes) results in a single potential barrier being replaced 
by a number of distributed barriers. 
It is now well known that a barrier
distribution can be extracted experimentally 
from the fusion excitation function 
$\sigma_{\rm fus}(E)$ by taking the second derivative of the product
$E\sigma_{\rm fus}(E)$ with respect to the center-of-mass energy $E$, that is, 
$d^2(E\sigma_{\rm fus})/dE^2$ \cite{RSS91}.  
The extracted fusion barrier distributions have been
found to be very sensitive to the structure of the colliding nuclei
\cite{DHRS98,L95}, 
and thus the barrier distribution method has opened up 
the possibility of 
exploiting the heavy-ion fusion reaction as a ``quantum tunneling
microscope'' in order to investigate both the static and 
dynamical properties of atomic nuclei. 

The same barrier distribution interpretation can be applied to 
the scattering process as well. 
In particular, it was suggested in Ref. \cite{ARN88} 
that the same information as the fusion cross
section may be obtained from the cross section for quasi-elastic
scattering (a sum of elastic, inelastic, and transfer cross sections) 
at large angles. 
Timmers {\it et al.}
proposed to use the first derivative of the ratio of the
quasi-elastic cross section $\sigma_{\rm qel}$ 
to the Rutherford cross section $\sigma_R$ with
respect to energy, $-d (d\sigma_{\rm qel}/d\sigma_R)/dE$, as an
alternative 
representation of the barrier distribution \cite{TLD95}. 
Their experimental data have revealed 
that the quasi-elastic barrier distribution is indeed similar to that
for fusion, although the former may be somewhat smeared and thus 
less sensitive to nuclear structure effects 
(see also Refs.\cite{PKP02,MSS03,SMO02} for recent measurements). 
As an example, we show in Fig. 1 a comparison between the
fusion and the quasi-elastic barrier distributions for the 
$^{16}$O + $^{154}$Sm system \cite{HR04}. 

\begin{figure}
\includegraphics[scale=0.4,clip]{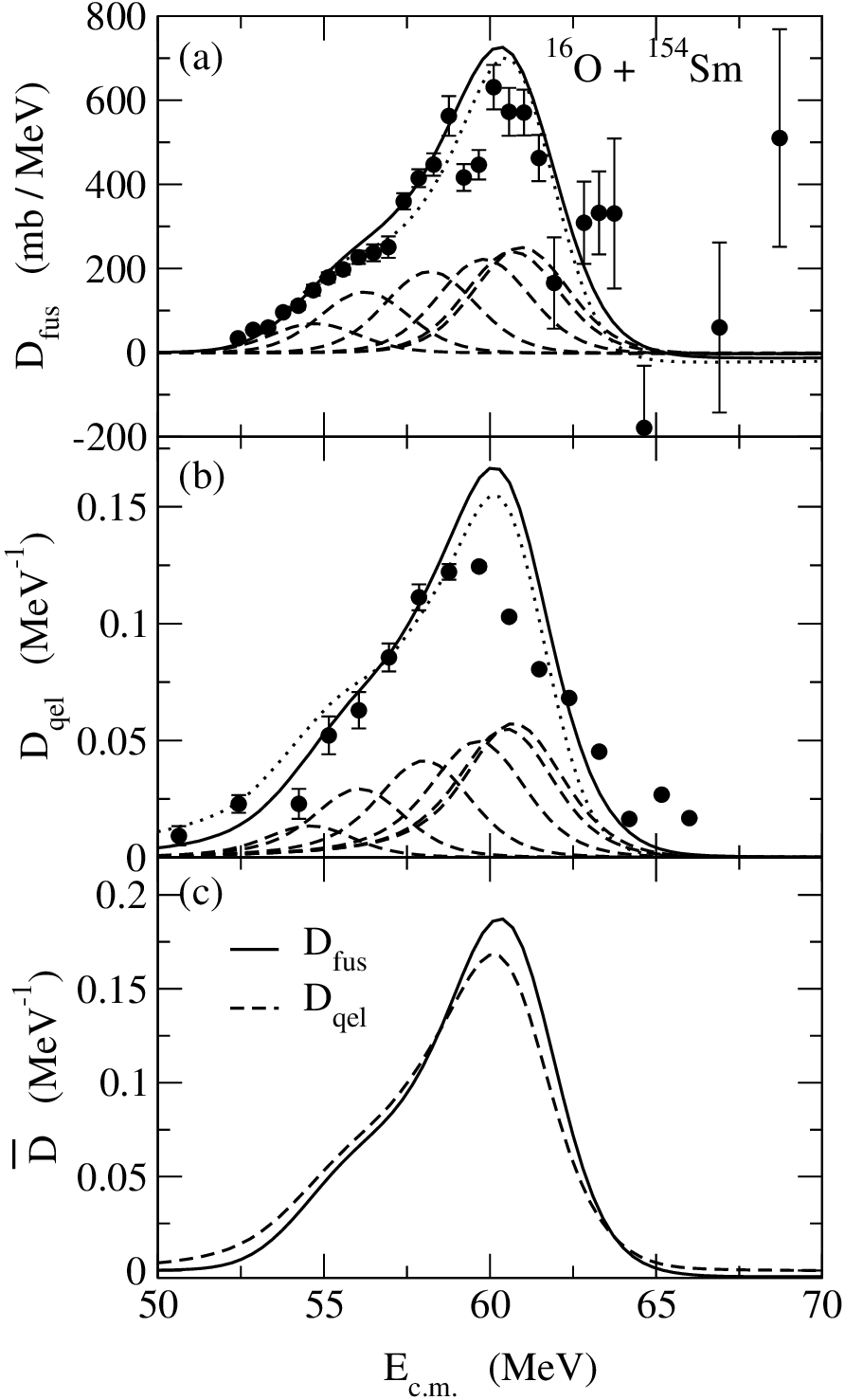}
\caption{
(a) The fusion barrier distribution 
for the $^{16}$O + $^{154}$Sm reaction. 
The solid line is obtained with the
orientation-integrated formula with $\beta_2=0.306$ and $\beta_4$=
0.05. 
The dashed lines indicate the contributions from the six
individual eigenbarriers. These lines are 
obtained by using a
Woods-Saxon potential with a surface diffuseness parameter $a$ of 0.65
fm. The dotted line is the fusion barrier distribution calculated with 
a potential which has $a$ = 1.05 fm. 
%Experimental data are taken from Ref. {\protect\cite{L95}}. 
(b) Same as Fig. 1(a), but for the quasi-elastic barrier
    distribution.  
%Experimental data are from Ref. {\protect\cite{TLD95}}. 
(c) Comparison between 
the barrier distribution for fusion (solid line)
    and that for quasi-elastic scattering (dashed line). 
These functions are both normalized to unit area in the energy interval 
between 50 and 70 MeV.}
\end{figure}

In this contribution, we 
undertake a detailed discussion of the 
properties of the quasi-elastic barrier distribution \cite{HR04}, 
which are less known than the fusion counterpart.
We shall discuss possible advantagges 
for its exploitation, putting a particular emphasis on future 
experiments with radioactive beams. 

\section{Quasi-elastic barrier distributions}

Let us first discuss 
heavy-ion reactions between inert nuclei. 
The classical fusion cross section is given by, 
\begin{equation}
\sigma^{cl}_{\rm fus}(E)=\pi
R_b^2\left(1-\frac{B}{E}\right)\,\theta(E-B),
\end{equation}
where $R_b$ and $B$ are the barrier position and the barrier height,
respectively. 
From this expression, it is clear that the first derivative of 
$E\sigma^{cl}_{\rm fus}$ is proportional to the classical 
penetrability for a 1-dimensional barrier of height $B$ or
eqivalently the s-wave penetrability, 
\begin{equation}
\frac{d}{dE}[E\sigma^{cl}_{\rm fus}(E)]=\pi R_b^2\,\theta(E-B)
=\pi R_b^2\,P_{cl}(E),
\end{equation}
and the second derivative to a delta function, 
\begin{equation}
\frac{d^2}{dE^2}[E\sigma^{cl}_{\rm fus}(E)]=\pi R_b^2\,\delta(E-B). 
\label{clfus}
\end{equation}

In quantum mechanics, 
the tunneling effect smears the delta function in Eq. (\ref{clfus}). 
If we define the fusion test function as
\begin{equation}
G_{\rm fus}(E)=\frac{1}{\pi R_b^2}\frac{d^2}{dE^2}
[E\sigma_{\rm fus}(E)], 
\end{equation}
this function has the following properties: 
i) it is symmetric around $E=B$, ii) it is centered on $E=B$, iii) 
its integral over $E$ is unity, and iv) it has a relatively narrow width
of around $\hbar\Omega\ln(3+\sqrt{8})/\pi \sim 0.56 \hbar\Omega$,
where $\hbar\Omega$ is the curvature of the Coulomb barrier. 

We next ask ourselves 
the question of how best to define a similar test function 
for a scattering problem. 
In the pure classical approach, in the limit of a strong Coulomb field, 
the differential cross sections for elastic scattering at $\theta=\pi$ 
is given by, 
\begin{equation}
\sigma_{\rm el}^{cl}(E,\pi)=\sigma_R(E,\pi)\,\theta(B-E), 
\end{equation}
where $\sigma_R(E,\pi)$ is the Rutherford cross section. 
Thus, the ratio $\sigma_{\rm el}^{cl}(E,\pi)/\sigma_R(E,\pi)$ is the
classical reflection probability $R(E)$ ($=1-P(E)$), and 
the appropriate test function for 
scattering is \cite{TLD95},
\begin{equation}
G_{\rm qel}(E)=-\frac{dR(E)}{dE} 
=-\frac{d}{dE}\left(\frac{\sigma_{\rm el}(E,\pi)}{\sigma_R(E,\pi)}\right). 
\label{qeltest}
\end{equation}

In realistic systems, due to the effect of nuclear
distortion, the differential cross section deviates from the
Rutherford cross section even at energies below the barrier. 
Using the semi-classical perturbation theory, 
we have derived 
a semi-classical formula for the backward scattering 
which takes into account the nuclear effect to the leading order \cite{HR04}. 
The result for a scattering angle $\theta$ 
reads, 
\begin{equation}
\frac{\sigma_{\rm el}(E,\theta)}{\sigma_R(E,\theta)}
=\alpha(E,\lambda_c)\cdot |S(E,\lambda_c)|^2,
\label{ratio}
\end{equation}
where 
$S(E,\lambda_c)$ is the total (Coulomb + nuclear)
$S$-matrix at energy $E$ and angular momentum 
$\lambda_c = \eta\cot(\theta/2)$, with $\eta$ being the usual Sommerfeld
parameter. 
Note that 
$|S(E,\lambda_c)|^2$ is nothing but the reflection probability of the
Coulomb barrier, $R(E)$. For $\theta=\pi$, $\lambda_c$ is zero, and 
$|S(E,\lambda_c=0)|^2$ is given by 
\begin{equation}
|S(E,\lambda_c=0)|^2 = R(E) = 
\frac{\exp\left[-\frac{2\pi}{\hbar\Omega}(E-B)\right]}
{1+\exp\left[-\frac{2\pi}{\hbar\Omega}(E-B)\right]} 
\end{equation}
in the parabolic approximation. 
$\alpha(E,\lambda_c)$ in Eq. (\ref{ratio}) is given by 
\begin{eqnarray}
\alpha(E,\lambda_c)&=&1+\frac{V_N(r_c)}{ka}\,
\frac{\sqrt{2a\pi k\eta}}{E}\,\\
&\times& 
\left[1-\frac{r_c}{Z_PZ_Te^2}\cdot
2V_N(r_c)
\left(\frac{r_c}{a}-1\right)\right],
\end{eqnarray}
where $k=\sqrt{2\mu E/\hbar^2}$, with $\mu$ being the reduced mass for
the colliding system. The nuclear potential $V_N(r_c)$ is evaluated at
the Coulomb turning point $r_c=(\eta+\sqrt{\eta^2+\lambda_c^2})/k$,
and $a$ is the diffuseness parameter in the nuclear potential. 

\begin{figure}
\includegraphics[scale=0.4,clip]{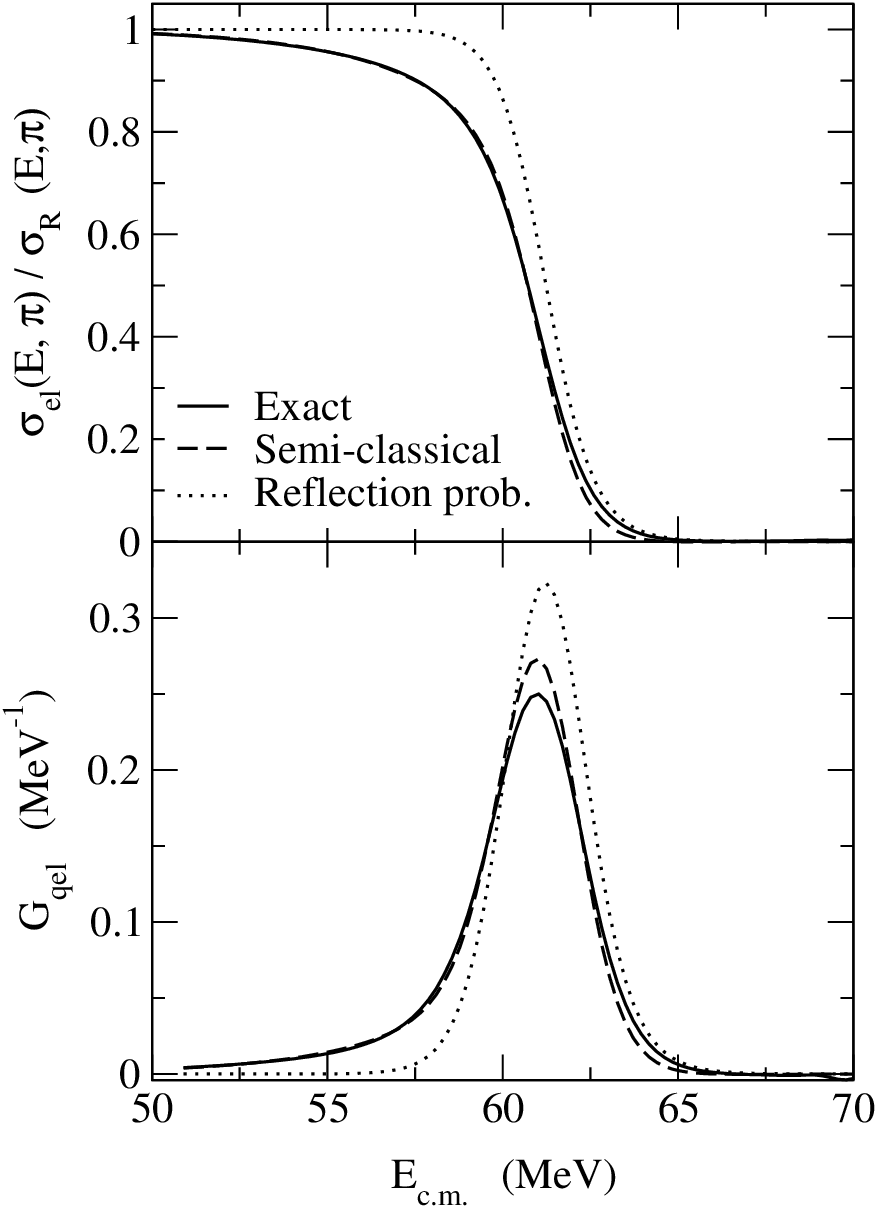}
\caption{
The ratio of elastic scattering to the Rutherford cross section at 
$\theta=\pi$ (upper panel) and the quasi-elastic test
function $G_{\rm qel}(E)=-d/dE (\sigma_{\rm el}/\sigma_R)$ (lower panel) 
for the $^{16}$O + $^{144}$Sm reaction. 
%The solid line is the exact solution of the optical potential, while
%the dashed line is obtained with the semi-classical perturbation
%theory. The dotted line denotes the reflection probability
%$R(E)=|S(E)|^2$ for $s$-wave scattering.
}
\end{figure}

Figure 2 shows an example of the 
excitation function of the cross sections 
and the corresponding quasi-elastic test function, $G_{\rm qel}$ 
at $\theta=\pi$ for the $^{16}$O + $^{144}$Sm reaction. 
Because of the nuclear distortion factor $\alpha(E,\lambda_c)$, 
the quasi-elastic test function behaves a little less simply 
than that for fusion. 
Nevertheless, the quasi-elastic test function 
$G_{\rm qel}(E)$ behaves rather similarly to the fusion test function 
$G_{\rm fus}(E)$. 
In particular, 
both functions have a similar, relatively narrow, width, 
and their integral over $E$ is unity. 
We may thus consider that the quasi-elastic test function is an excellent 
analogue of the one for fusion, and we exploit this fact in 
studying barrier structures in heavy-ion scattering. 

In the presence of the channel couplings, 
the fusion and the quasi-elastic 
cross sections may be given as a weighted sum of the cross sections 
for uncoupled eigenchannels, 
\begin{eqnarray}
\sigma_{\rm fus}(E)&=&\sum_\alpha w_\alpha 
\sigma_{\rm fus}^{(\alpha)}(E), \label{crossfus}\\
\sigma_{\rm qel}(E,\theta)&=&\sum_\alpha w_\alpha 
\sigma_{\rm el}^{(\alpha)}(E,\theta), \label{crossqel}
\end{eqnarray}
where $\sigma_{\rm fus}^{(\alpha)}(E)$ and 
$\sigma_{\rm el}^{(\alpha)}(E,\theta)$
are the fusion and the elastic cross sections for a potential 
in the eigenchannel $\alpha$. 
These equations 
immediately lead to 
the expressions for the barrier distribution in terms of the test
functions, 
\begin{eqnarray}
D_{\rm fus}(E)&=&\frac{d^2}{dE^2}[E\sigma_{\rm fus}(E)]=
\sum_\alpha w_\alpha 
\pi R_{b,\alpha}^2\,G_{\rm fus}^{(\alpha)}(E), 
\label{weightedsum}
\\
D_{\rm qel}(E)&=&
-\frac{d}{dE}\left(\frac{\sigma_{\rm qel}(E,\pi)}{\sigma_R(E,\pi)}\right) 
=
\sum_\alpha w_\alpha 
G_{\rm qel}^{(\alpha)}(E). 
\end{eqnarray}

\section{Advantages over fusion barrier distributions}

There are certain attractive experimental advantages to 
measuring the quasi-elastic 
cross section $\sigma_{\rm qel}$ rather than the fusion cross sections 
$\sigma_{\rm fus}$ to extract a representation of 
the barrier distribution. 
These are: i) less accuracy is required in the data for taking the
first derivative rather than the second derivative, ii) whereas
measuring the fusion cross section requires specialized recoil
separators (electrostatic deflector/velocity filter) usually of low
acceptance and efficiency, the measurement of 
$\sigma_{\rm qel}$ needs only very simple charged-particle detectors,
not necessarily possessing good resolution either in energy or in
charge, and iii) several effective energies can be measured at a
single-beam energy, since, in the semi-classical approximation, each 
scattering angle corresponds to scattering at a certain angular
momentum, and the cross section can be scaled in energy by taking 
into account the centrifugal correction. 
Estimating the centrifugal potential at the Coulomb turning 
point $r_c$, the effective energy may be expressed as \cite{TLD95}
\begin{equation}
E_{\rm eff}\sim E
-\frac{\lambda_c^2\hbar^2}{2\mu r_c^2} 
=2E\frac{\sin(\theta/2)}{1+\sin(\theta/2)}.
\label{Eeff}
\end{equation}
Therefore, one expects that the function 
$-d/dE (\sigma_{\rm el}/\sigma_R)$ 
evaluated 
at an angle $\theta$ will correspond to the quasi-elastic test function 
(\ref{qeltest}) 
at the effective energy given by eq. (\ref{Eeff}). 

This last point not only improves the efficiency of the experiment, 
but also allows the use of a cyclotron accelerator where the
relatively small energy steps required for barrier 
distribution experiments cannot be obtained from the
machine itself \cite{PKP02}. 
Moreover, these advantages all point to greater ease of measurement with
low-intensity exotic beams, which will be discussed in the 
next section. 

\begin{figure}
\includegraphics[scale=0.4,clip]{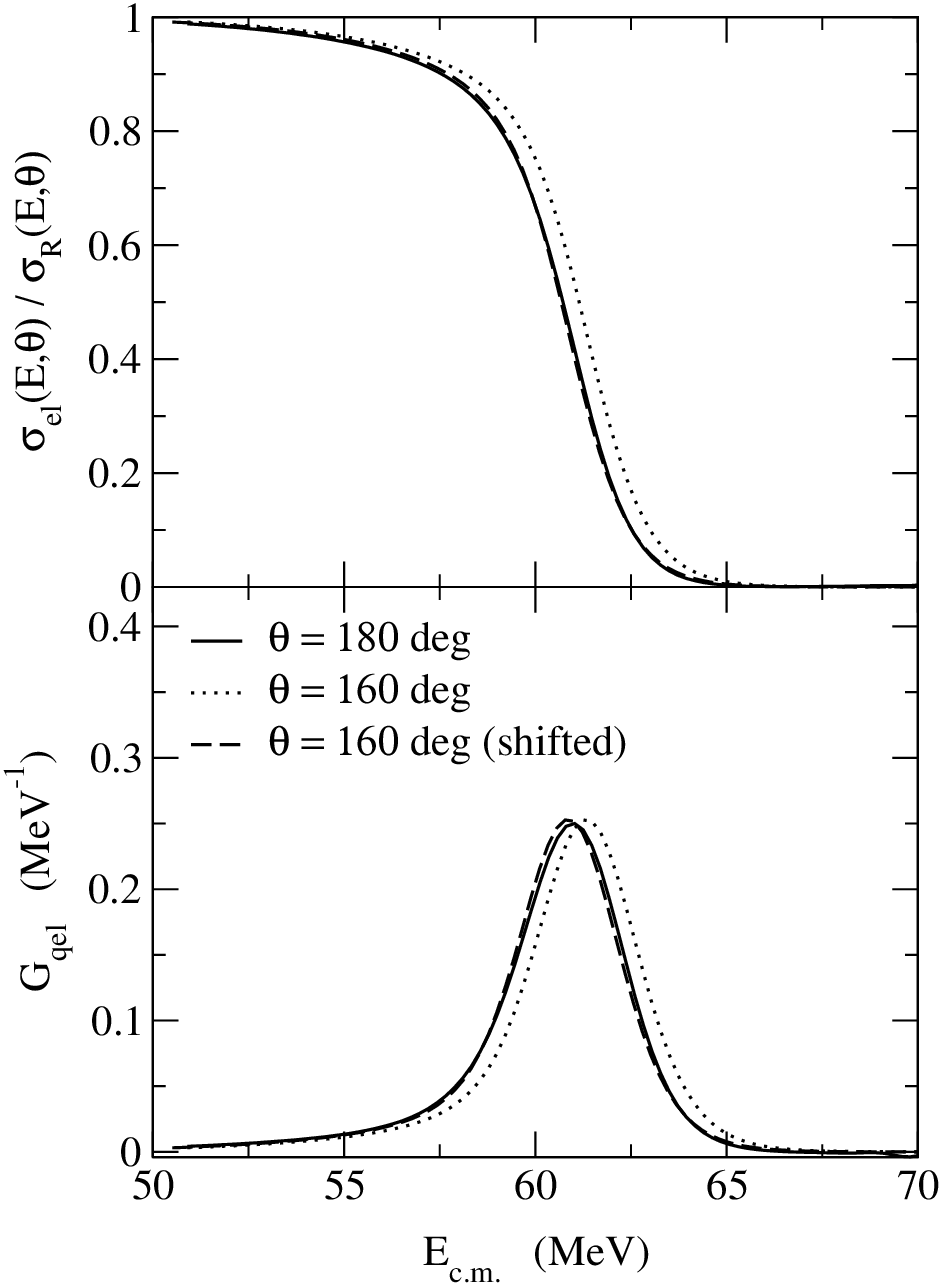}
\caption{
Comparison of the 
ratio $\sigma_{\rm el}/\sigma_R$ (upper panel) and its energy 
derivative $-d/dE (\sigma_{\rm el}/\sigma_R)$ (lower panel) 
evaluated at two 
different scattering angles. 
%The solid line is for $\theta=\pi$, while 
%the dotted line is for $\theta=160^{\rm o}$. 
%The dashed line is the same as the dotted line, but is shifted in energy 
%by an amount equal to 
%the centrifugal potential evaluated at the distance of closest
%approach of the 
%Rutherford trajectory.
}
\end{figure}

In order to check the scaling property of the quasi-elastic test function 
with respect to the angular momentum, 
Fig. 3 compares the functions  
$\sigma_{\rm el}/\sigma_R$ (upper panel) and 
$-d/dE (\sigma_{\rm el}/\sigma_R)$ (lower panel) 
obtained at two different scattering angles. 
The solid line is evaluated at $\theta=\pi$, while the dotted line at 
$\theta=160^{\rm o}$. The dashed line is the same as the dotted line, but 
shifted in energy by $E_{\rm eff}-E$. 
Evidently, the scaling does work well, both at 
energies below and above the Coulomb barrier, although 
it becomes less good 
as the scattering angle decreases \cite{HR04}. 

\section{Quasi-elastic scattering with radioactive beams}

Low-energy radioactive beams have become increasingly 
available in recent years, 
and heavy-ion fusion reactions involving neutron-rich nuclei 
have been performed for a few systems 
\cite{SYW04,LSG03,RSC04}. 
New generation facilities have been under 
construction at several laboratories, and many more 
reaction measurements 
with exotic beams at low energies will be performed in the near future. 
Although it would still be difficult to perform high-precision 
measurements of fusion cross sections with radioactive beams, 
the measurement of the quasi-elastic barrier distribution, which 
can be obtained much more easily than the fusion 
counterpart as we discussed in the previous section, may be feasible.
Since the quasi-elastic barrier distribution contains similar
information as the fusion barrier distribution, 
the quasi-elastic measurements at backward angles  
may open up a 
novel way to probe the structure of exotic neutron-rich nuclei. 

\begin{figure}
\includegraphics[scale=0.4,clip]{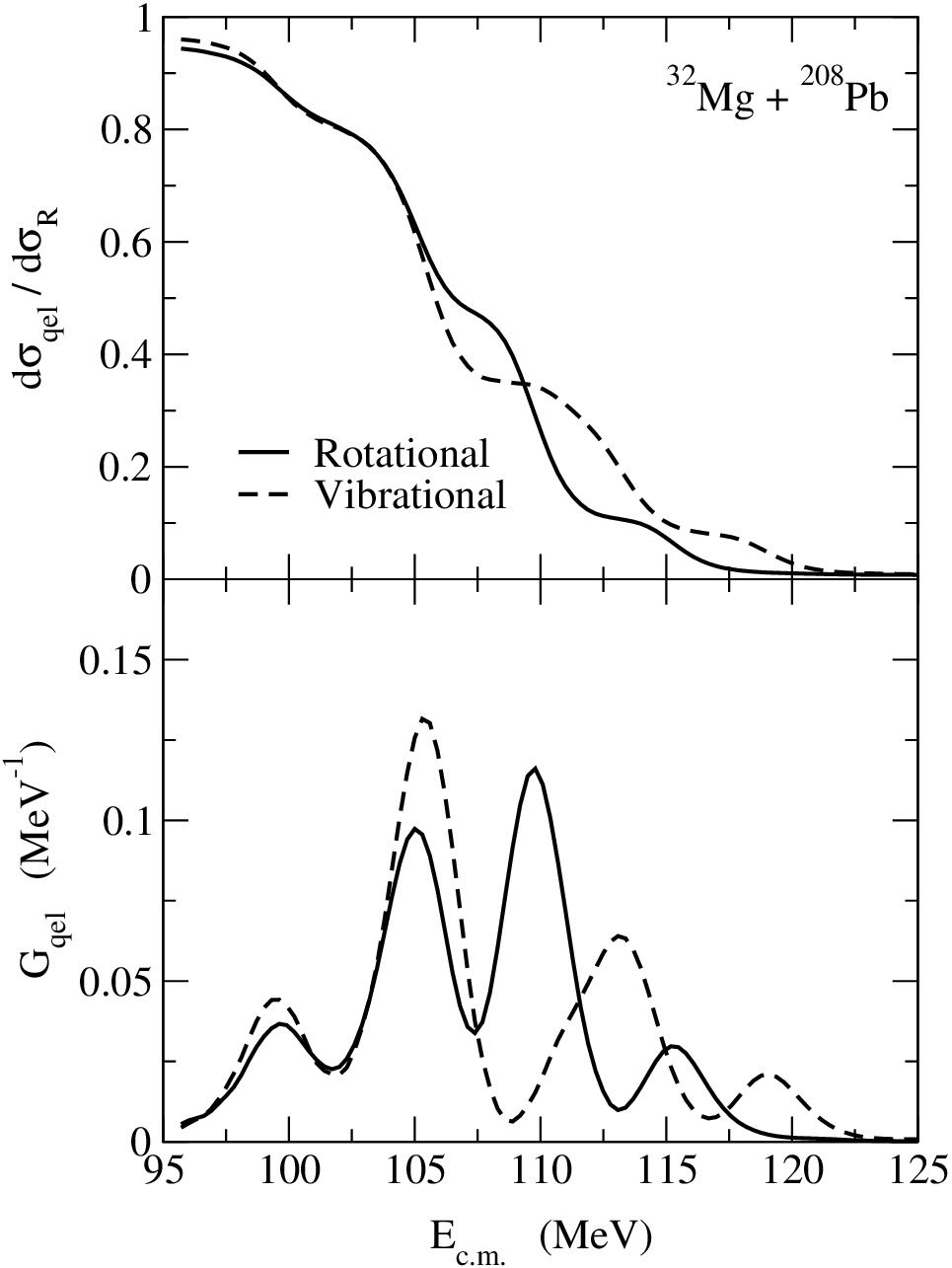}
\caption{
The excitation function for quasi-elastic scattering  
(upper panel) and the quasi-elastic barrier distribution 
(lower panel) for the 
$^{32}$Mg + $^{208}$Pb reaction around the Coulomb barrier. 
The solid and the dashed lines are the results of coupled-channels
calculations which assume that $^{32}$Mg is a rotational and a
vibrational nucleus, respectively. 
The single octupole-phonon excitation in 
$^{208}$Pb is also included in the calculations. }
\end{figure}

In order to demonstrate the usefulness of the study of the 
quasi-elastic barrier
distribution with radioactive beams, we take as an example the
reaction $^{32}$Mg and $^{208}$Pb, where the quadrupole 
collectivity of the neutron-rich $^{32}$Mg remains to be clarified 
experimentally. 
Fig. 4 shows the excitation function of the quasi-elastic scattering
(upper panel) and the quasi-elastic barrier distribution 
(lower panel) for this system. 
The solid and dashed lines are results of coupled-channels
calculations where $^{32}$Mg is assumed to be a rotational or
a vibrational nucleus, respectively. 
We include the quadrupole excitations in $^{32}$Mg up to the second 
member (that is, the first 4$^+$ state in the rotational band for the
rotational coupling, or the double phonon state for the vibrational
coupling). In addition, we include the single octupole phonon
excitation at 2.615 MeV in $^{208}$Pb. 
We use a version of the computer code {\tt CCFULL} \cite{HRK99} in order 
to integrate the
coupled-channels equations. 
One clearly sees well separated peaks in the
quasi-elastic barrier distribution both for the rotational and for the
vibrational couplings. Moreover, the two lines are considerably
different at energies around and above the Coulomb barrier, although 
the two results are rather similar below the
barrier. 
We can thus expect that the
quasi-elastic barrier distribution can indeed be utilized 
to discriminate between the rotational and the
vibrational nature of the quadrupole collectivity in $^{32}$Mg, 
although these results might be somewhat perturbed by other
effects which are not considered in the present calculations, such as
double octupole-phonon excitations in the target, transfer
processes or hexadecapole deformations. 

We mention that the distorted-wave Born approximation (DWBA) yields 
identical results for both
rotational and vibrational couplings (to first order). 
In order to discriminate whether the transitions are vibration-like
or rotation-like, at least second-step processes (reorientation
and/or couplings to higher members) are necessary. 
The coupling effect plays a more important role in low-energy
reactions than at high and intermediate energies. Therefore, 
we expect that quasi-elastic scattering around the 
Coulomb barrier will provide a useful means 
to allow 
the detailed study of the structure of neutron-rich nuclei in 
the near future. 

\begin{acknowledgments}
This work was supported by the Grant-in-Aid for Scientific Research,
Contract No. 16740139, 
from the Japan Society for the Promotions of
Science. 
\end{acknowledgments}


\begin{thebibliography}{99}

\bibitem{DHRS98}M. Dasgupta, D.J. Hinde, N. Rowley, and 
A.M. Stefanini, Annu. Rev. Nucl. Part. Sci. {\bf 48}, 401 (1998). 

\bibitem{BT98}
A.B. Balantekin and N. Takigawa,
Rev. Mod. Phys. {\bf 70}, 77 (1998). 

\bibitem{RSS91}N. Rowley, G.R. Satchler, and P.H. Stelson, 
Phys. Lett. {\bf B254}, 25 (1991). 

\bibitem{L95}J.R. Leigh {\it et al.}, 
Phys. Rev. C{\bf 52}, 3151 (1995). 

\bibitem{ARN88}M.V. Andres, N. Rowley, and M.A. Nagarajan, 
Phys. Lett. {\bf 202B}, 292 (1988). 

\bibitem{TLD95}H. Timmers {\it et al.}, 
Nucl. Phys. {\bf A584}, 190 (1995). 

\bibitem{PKP02}E. Piasecki {\it et al.}, 
Phys. Rev. C{\bf 65}, 054611 (2002). 

\bibitem{MSS03}D.S. Monteiro {\it et al.}, Nucl. Phys. {\bf A725}, 60
  (2003). 

\bibitem{SMO02}R.F. Simoes {\it et al.}, Phys. Lett. {\bf B527}, 
187 (2002). 

\bibitem{HR04}K. Hagino and N. Rowley, Phys. Rev. C{\bf 69}, 
054610 (2004). 

\bibitem{SYW04}C. Signorini {\it et al.}, 
Nucl. Phys. {\bf A735}, 329 (2004). 

\bibitem{LSG03}J.F. Liang {\it et al.}, 
Phys. Rev. Lett. {\bf 91}, 152701 (2003). 

\bibitem{RSC04}R. Raabe {\it et al.}, Nature {\bf 431}, 823 (2004). 

\bibitem{HRK99}K. Hagino, N. Rowley, and A.T. Kruppa,
Comp. Phys. Comm. {\bf 123}, 143 (1999). 

\end{thebibliography}
\end{document}